\documentclass[12pt,english]{article}
\usepackage[T1]{fontenc}
\usepackage[latin9]{inputenc}
\usepackage[a4paper]{geometry}
\usepackage{amsmath}

\makeatletter
\newcommand{\lyxaddress}[1]{
	\par {\raggedright #1
	\vspace{1.4em}
	\noindent\par}
}

\makeatother

\usepackage{babel}
\begin{document}
\title{$W-$exchange contributions in hadronic decays of bottom baryon $\Lambda_{b}$}
\author{Fayyazuddin}
\maketitle

\lyxaddress{Physics Department , Quaid-i-Azam University, Islamabad 45320, Pakistan.}
\begin{abstract}
The nonleptonic decays $\Lambda_{b}\rightarrow\Sigma_{c}^{*-}\pi^{+},\Xi_{c}^{*0}K^{0}$
and $\Lambda_{b}\rightarrow\Delta^{0}D^{0},\Sigma^{*-}D_{0}^{+}$
are studied. In addition, the decays $\Lambda_{b}\rightarrow\Xi_{c}^{0}K^{0},\Sigma^{-}D_{s}^{+}$
are analyzed. For all these decays the dominant contribution comes
from $W-$exchange, and for the decay $\Lambda_{b}\rightarrow\Lambda_{c}^{+}\pi^{-}$,
in addition to factorization, baryon pole contribution to the $p$-wave
(parity conserving) decay amplitude $B$ is discussed.
\end{abstract}
In this paper, the formalism developed in \cite{key-1} based on \cite{key-2,key-3}
for the charm baryons is closely followed for the bottom baryons.

Since $\mathcal{B}_{b}^{\prime}\left(\frac{1}{2}^{+}\right)$ belongs
to the triplet representation $\bar{3}$ of $SU(3)$, The matrix elements 

\[
\left\langle \mathcal{B}^{*}\left(\frac{3}{2}^{+}\right),\mathcal{B}_{c}^{*}\left(\frac{3}{2}^{+}\right)\right|\bar{q}\gamma_{\mu}(1-\gamma_{5})b\left|\mathcal{B}_{b}^{\prime}\left(\frac{1}{2}^{+}\right)\right\rangle =0
\]
where $q=u,d,s$. Thus, the factorization does not contribute to the
decays: $\mathcal{B}_{b}^{\prime}\left(\frac{1}{2}^{+}\right)\rightarrow\mathcal{B}_{c}^{*}\left(\frac{3}{2}^{+}\right)+P$
and $\mathcal{B}_{b}^{\prime}\left(\frac{1}{2}^{+}\right)\rightarrow\mathcal{B}^{*}\left(\frac{3}{2}^{+}\right)+P_{D}.$
Hence the dominant contribution comes from the $W-$exchange.

$W-$exchange in the non-relativistic limit is encoded in the effective
Hamiltonian

\begin{equation}
H_{W}^{PC}=\frac{G_{F}}{\sqrt{2}}V_{cb}V_{ud}\sum_{i\ne j}\alpha_{i}^{-}\gamma_{j}^{+}(1-\sigma_{i}\cdot\sigma_{j})\delta^{3}(r),\label{eq:1}
\end{equation}
where 
\begin{align*}
\alpha_{i}^{-}\left|u\right\rangle  & =\left|d\right\rangle ,\:i=1,2\:\text{to be summed over }i\\
\gamma_{j}^{\dagger}\left|b\right\rangle  & =\left|c\right\rangle ,\:j=3.
\end{align*}
$W-$exchange is relevant when one consider the baryon pole contribution
(Born terms) involving matrix elements of form $\left\langle \mathcal{B}_{c}\right|H_{W}^{PC}\left|\mathcal{B}_{b}\right\rangle $
which can be evaluated in the non-relativistic quark model (NQM) by
using Eq. (\ref{eq:1})

One notes:

\begin{equation}
[\alpha_{1}^{-}\gamma_{3}^{+}(1-\sigma_{1}.\sigma_{3})+\alpha_{2}^{-}\gamma_{3}^{+}(1-\sigma_{2}.\sigma_{3})]\left|\Lambda_{b}\right\rangle =\sqrt{6}\left|\Sigma_{c}^{0}\right\rangle .\label{eq:2}
\end{equation}
Here, $\left|\Sigma_{c}^{0}\right\rangle $ belong to symmetric representation
$6$ of $SU(3)$,

\begin{align*}
\bar{3}\times8 & =6+15+\bar{3},\\
\bar{3}\times10 & =6+24.
\end{align*}
Hence, the dominant contribution to the decays 

\begin{align*}
\Lambda_{b} & \rightarrow\Sigma_{c}^{*+}\pi^{-},\Xi_{c}^{*0}K^{0},\\
\Lambda_{b} & \rightarrow\Delta^{0}D^{0},\Sigma^{*-}D_{s}^{+},
\end{align*}
comes from the $\Sigma_{c}^{0}$ pole, i.e., from the chain

\begin{equation}
\Lambda_{b}\rightarrow\Sigma_{c}^{0}\rightarrow\Sigma_{c}^{*+}\pi^{-},\Xi_{c}^{*0}K^{0},\label{eq:3}
\end{equation}

\begin{equation}
\Lambda_{b}\rightarrow\Sigma_{c}^{0}\rightarrow\Delta^{0}D^{0},\Sigma^{*-}D_{s}^{+}.\label{eq:4}
\end{equation}
The decay rate for the decays of the form $B_{b}(b^{\prime})\rightarrow B_{c}^{*}(p)+P(q)$
is given by \cite{key-1}:

\begin{equation}
\Gamma=\frac{1}{6\pi}\frac{m^{\prime}}{m^{*}}\frac{\left|\vec{p}\right|^{3}}{f_{p}^{2}}\left(p_{0}+m^{*}\right)\left|C\right|^{2}.\label{eq:5}
\end{equation}
From Eq. (\ref{eq:5}), we have

\begin{equation}
\Gamma\left(\Lambda_{b}\rightarrow\Sigma_{c}^{*+}\pi^{-}\right)\approx2.14\times10^{2}\left|C\right|^{2},\label{eq:6}
\end{equation}

\begin{equation}
\Gamma\left(\Lambda_{b}\rightarrow\Xi_{c}^{*0}K^{0}\right)\approx1.00\times10^{2}\left|C\right|^{2},\label{eq:7}
\end{equation}
and 
\begin{equation}
\Gamma\left(\Lambda_{b}\rightarrow\Delta^{0}D^{0}\right)\approx3.99\times10^{2}\left|C\right|^{2},\label{eq:8}
\end{equation}

\begin{equation}
\Gamma\left(\Lambda_{b}\rightarrow\Sigma^{*-}D_{s}^{+}\right)\approx1.31\times10^{2}\left|C\right|^{2}.\label{eq:9}
\end{equation}

In order to determine the amplitude C, one notes

\begin{align}
S_{ij}^{*}P_{k}^{j}S^{ik} & \rightarrow\left[S_{2j}^{*}P_{2}^{j}\right]S^{22},\nonumber \\
 & =\left[\Sigma_{c}^{*+}\pi^{-}-\Sigma_{c}^{*0}\pi^{0}+\Xi_{c}^{*0}K^{0}\right]\sqrt{2}\Sigma_{c}^{0},\label{eq:10}
\end{align}
and

\begin{align}
F_{i}T_{jk}S^{jk} & \rightarrow\left[F_{i}T_{i22}\right]S^{22},\nonumber \\
 & =\left[D^{0}T_{122}+D^{+}T_{222}+D_{s}^{*}T_{322}\right]\sqrt{2}\Sigma_{c}^{0},\nonumber \\
 & =\left[\Delta^{0}D^{0}+\sqrt{3}\Delta^{-}D^{+}+\Sigma^{*-}D_{s}^{+}\right]2\Sigma_{c}^{0}.\label{eq:11}
\end{align}
The above results are in agreement with the one derived in \cite{key-4}.

The amplitude $C$ for the decays $\Lambda_{b}\rightarrow\Sigma_{c}^{*+}\pi^{-},\:\Xi_{c}^{*0}K^{0}$
is given by \cite{key-1}.

\begin{equation}
C=(1,1)\sqrt{2}g_{c}^{*}\frac{\left\langle \Sigma_{c}^{0}\right|H_{W}^{PC}\left|\Lambda_{b}\right\rangle }{m_{\Lambda_{b}}-m_{\Sigma_{c}^{0}}},\label{eq:12}
\end{equation}
and for the decays $\Lambda_{b}\rightarrow\Delta^{0}D^{0},\Sigma^{*-}D_{s}^{*-}$,

\begin{equation}
C=(1,1)2g_{c}^{*}\frac{\left\langle \Sigma_{c}^{0}\right|H_{W}^{PC}\left|\Lambda_{b}\right\rangle }{m_{\Lambda_{b}}-m_{\Sigma_{c}^{0}}}.\label{eq:13}
\end{equation}
The weak matrix elements $\left\langle \Sigma_{c}^{0}\right|H_{W}^{PC}\left|\Lambda_{b}\right\rangle $
on using Eq. (\ref{eq:1}) is given by 

\begin{equation}
\left\langle \Sigma_{c}^{0}\right|H_{W}^{PC}\left|\Lambda_{b}\right\rangle =\left[\frac{G_{F}}{\sqrt{2}}V_{cb}V_{ud}\right]\sqrt{6}d^{\prime}.\label{eq:14}
\end{equation}

In order to determine $d^{\prime}$, one notes that the decays $S^{*}\rightarrow S\pi$
is not energetically allowed. However the decays $S^{*}\rightarrow A\pi$
is possible and the decay rate is given by 

\begin{equation}
\Gamma(S^{*}\rightarrow A\pi)=\frac{1}{12\pi}\frac{1}{m_{S^{*}}}\frac{\left(\sqrt{2}g_{c}^{*}\right)^{2}}{f_{\pi}^{2}}\left(p_{0}+m_{A}\right)\left|\vec{p}\right|^{3}.\label{eq:15}
\end{equation}
$SU(3)$ gives

\[
\sqrt{2}g_{c}^{*}\left[\Sigma_{c}^{*++}\pi^{+}-\Sigma_{c}^{*+}\pi^{0}\right]\Lambda_{c}^{+},
\]
from Eq. (\ref{eq:15})

\begin{equation}
\Gamma(\Sigma_{c}^{*++}\rightarrow\Lambda_{c}^{*}\pi^{+})=\left(\sqrt{2}g_{c}^{*}\right)^{2}(1.673).\label{eq:16}
\end{equation}
From the experimental value for the decay rate, we get

\begin{equation}
\sqrt{2}g_{c}^{*}\approx0.94.\label{eq:17}
\end{equation}
Hence \cite{key-5}, after using the following numerical values

\[
\alpha_{s}=0.32,\:m_{\Sigma_{c}^{*}}-m_{\Lambda_{c}}=0.231\:\text{GeV},\:m_{b}=4.85\,\text{GeV}\:
\]

\[
m_{d}=m_{u}=0.336\,\text{GeV},\,m_{s}=0.510\,\text{GeV}
\]

\[
m_{c}=1.45\:\text{GeV},\:m_{bd}=\frac{m_{b}m_{d}}{m_{b}+m_{d}}=0.313\:\text{GeV},\:m_{cu}=0.273\:\text{GeV},
\]
we have

\begin{equation}
d^{\prime}=\frac{3\left(m_{\Sigma_{c}^{*}}-m_{\Lambda_{c}}\right)}{8\pi\alpha_{s}}\left(m_{bd}\;m_{cu}\right)\approx7.4\times10^{-3}\,\text{GeV}{}^{3}.\label{eq:18}
\end{equation}

From Eqs. (\ref{eq:14}, \ref{eq:18})

\begin{equation}
\left\langle \Sigma_{c}^{0}\right|H_{W}^{PC}\left|\Lambda_{b}\right\rangle \approx7.4\times10^{-9}\:\text{GeV}.\label{eq:19}
\end{equation}
Hence, from Eqs. (\ref{eq:12}, \ref{eq:17}) and Eq. (\ref{eq:19}),
we have

\begin{equation}
C=(1,1)\sqrt{2}g_{c}^{*}\frac{\left\langle \Sigma_{c}^{0}\right|H_{W}^{PC}\left|\Lambda_{b}\right\rangle }{m_{\Lambda_{b}}-m_{\Sigma_{c}^{0}}}=(1,1)\left(2.2\times10^{-9}\right),\label{eq:20}
\end{equation}
for the decays $\Lambda_{b}\rightarrow\Sigma_{c}^{*+}\pi^{-},\Xi_{c}^{*0}K^{0}$,
and from (\ref{eq:13}, \ref{eq:17}) and Eq. (\ref{eq:19}):

\begin{equation}
C=(1,1)2g_{c}^{*}\frac{\left\langle \Sigma_{c}^{0}\right|H_{W}^{PC}\left|\Lambda_{b}\right\rangle }{m_{\Lambda_{b}}-m_{\Sigma_{c}^{0}}}=(1,1)\left(3.1\times10^{-9}\right),\label{eq:21}
\end{equation}
for the decays $\Lambda_{b}\rightarrow\Delta^{0}D^{0},\Sigma^{*-}D_{s}^{*}$.
From Eqs. (\ref{eq:6},\ref{eq:7}) and (\ref{eq:20}):

\begin{equation}
B_{r}(\Lambda_{b}\rightarrow\Sigma_{c}^{*+}\pi^{-})\approx2.9\times10^{-3},\label{eq:22}
\end{equation}

\begin{equation}
B_{r}(\Lambda_{b}\rightarrow\Xi_{c}^{*0}K^{0})\approx1.1\times10^{-3}.\label{eq:23}
\end{equation}

To take into account the $SU(3)$ symmetry breaking, the branching
ratio in Eq. (\ref{eq:23}) is multiplied by a factor $\frac{m_{s}^{2}}{m_{bd}m_{cu}}\approx3.0$,
and with this modification 

\begin{equation}
B_{r}(\Lambda_{b}\rightarrow\Xi_{c}^{*0}K^{0})\approx3.3\times10^{-3}.\label{eq:24}
\end{equation}

From Eqs (\ref{eq:8},\ref{eq:9}) and (\ref{eq:21}): 

\begin{equation}
B_{r}(\Lambda_{b}\rightarrow\Delta^{0}D^{0})\approx8.4\times10^{-3},\label{eq:25}
\end{equation}

\begin{equation}
B_{r}(\Lambda_{b}\rightarrow\Sigma^{*-}D_{s}^{+})\approx2.9\times10^{-3},\label{eq:26}
\end{equation}

\begin{equation}
B_{r}(\Lambda_{b}\rightarrow\Delta^{-}D^{+})\approx2.5\times10^{-2}.\label{eq:27}
\end{equation}

The experimental data is not available to check the result given in
Eqs. (\ref{eq:22}, \ref{eq:24}) and Eqs. (\ref{eq:25}, \ref{eq:26},
\ref{eq:27}).

There is nothing wrong in the basic formalism used in \cite{key-4},
but the parameters used to obtain the results are not realistic and
also there are other error. Hence the results obtained in reference
\cite{key-4} are unreliable and should be discarded.

Another aspect of this paper is the analyze the non-leptonic decays
of $\Lambda_{b}$:

\[
\Lambda_{b}(p^{\prime})\rightarrow\Lambda_{c}^{+}\pi^{-},\Xi_{c}^{0}K^{0},\Sigma^{-}D_{s}^{+},\quad p^{\prime}=p+k
\]
For the decays $\Lambda_{b}\rightarrow\Xi_{c}^{0}K^{0},\Sigma^{-}D_{s}^{+},$
factorization does not contribute; baryon pole contribution is the
dominant contribution to the $p-$wave (parity conserving) amplitude
$B=B_{\text{pole}}$. However for the decay $\Lambda_{b}\rightarrow\Lambda_{c}^{+}\pi^{-},$
factorization contributes. Hence the decay amplitude: 

\[
B=B_{\text{factorization}}+B_{\text{pole}},
\]

\begin{equation}
B_{\text{pole}}=g_{\Lambda_{c}\pi^{-}\Sigma_{Q}^{+}}\frac{1}{m_{\Sigma_{Q}^{*}}-m_{\Lambda_{c}}}\left\langle \Lambda_{c}^{*}\right|H_{W}^{PC}\left|\Sigma_{Q}^{+}\right\rangle .\label{eq:28}
\end{equation}
For the decay $\Lambda_{b}\rightarrow\Xi_{c}^{0}K^{0}$:

\begin{equation}
B=B_{\text{pole}}=g_{\Lambda_{b}K^{0}\Xi_{b}^{\prime0}}\frac{1}{m_{\Xi_{b}^{0\prime}}-m_{\Xi_{c}^{0}}}\left\langle \Xi_{c}^{0}\right|H_{W}^{PC}\left|\Xi_{b}^{0\prime}\right\rangle ,\label{eq:29}
\end{equation}
and for the decay $\Lambda_{b}\rightarrow\Sigma^{-}D_{s}^{+};$

\begin{equation}
B=B_{\text{pole}}=g_{\Lambda_{b}D_{s}^{+}\Sigma_{b}^{-}}\frac{1}{m_{\Sigma_{b}^{-}}-m_{\Sigma^{-}}}\left\langle \Sigma^{-}\right|H_{W}^{PC}\left|\Sigma_{b}^{-}\right\rangle .\label{eq:30}
\end{equation}

Now

\begin{equation}
\left\langle \Lambda_{c}^{*}\right|H_{W}^{PC}\left|\Sigma_{b}^{+}\right\rangle =\left\langle \Xi_{c}^{0}\right|H_{W}^{PC}\left|\Xi_{b}^{0\prime}\right\rangle =\left\langle \Sigma^{-}\right|H_{W}^{PC}\left|\Sigma_{b}^{-}\right\rangle =\left[\frac{G_{F}}{\sqrt{2}}V_{cb}V_{ud}\right]\sqrt{6}d^{\prime}\label{eq:31}
\end{equation}

The Goldberger-Treiman relation gives

\begin{equation}
g_{\Lambda_{b}\pi^{-}\Sigma_{b}^{+}}=\frac{m_{\Lambda_{b}}+m_{\Sigma_{b}^{+}}}{f_{\pi}}g_{A},\label{eq:32}
\end{equation}

\begin{equation}
g_{\Lambda_{b}K^{0}\Xi_{b}^{\prime0}}=\frac{m_{\Lambda_{b}}+m_{\Xi_{b}^{\prime0}}}{f_{K}}g_{A}^{\prime},\label{eq:33}
\end{equation}

\begin{equation}
g_{\Lambda_{b}D_{s}^{+}\Sigma_{b}^{-}}=\frac{m_{\Lambda_{b}}+m_{\Sigma_{b}^{-}}}{f_{D_{s}}}g_{A}^{\prime\prime}.\label{eq:34}
\end{equation}
The quark model gives 

\begin{equation}
g_{A}=\sqrt{\frac{2}{3}},g_{A}^{\prime}=\frac{1}{\sqrt{3}},g_{A}^{\prime\prime}=-\sqrt{\frac{2}{3}}.\label{eq:35}
\end{equation}
Hence for the decay $\Lambda_{b}\rightarrow\Lambda_{c}^{+}\pi^{-}$;
from Eqs. (\ref{eq:28}, \ref{eq:31}, \ref{eq:32}) and Eq. (\ref{eq:35}),
we have

\begin{equation}
B_{pole}=\left[\frac{G_{F}}{\sqrt{2}}V_{cb}V_{ud}\right]\left(\frac{m_{\Lambda_{b}^{+}}+m_{\Sigma_{b}^{+}}}{m_{\Sigma_{b}^{+}}-m_{\Lambda_{c}^{+}}}\right)\sqrt{6}d^{\prime}\frac{\sqrt{\frac{2}{3}}}{f_{\pi}}.\label{eq:36}
\end{equation}

For the decay $\Lambda_{b}\rightarrow\Xi_{c}^{0}K^{0};$ from Eqs.
(\ref{eq:29}, \ref{eq:31}, \ref{eq:33}) and Eq. (\ref{eq:35})

\begin{equation}
B=B_{pole}=\left[\frac{G_{F}}{\sqrt{2}}V_{cb}V_{ud}\right]\left(\frac{m_{\Lambda_{b}^{+}}+m_{\Xi_{b}^{\prime0}}}{m_{\Xi_{b}^{\prime0}}-m_{\Xi_{c}^{0}}}\right)\sqrt{6}d^{\prime}\frac{\sqrt{\frac{1}{3}}}{f_{K}},\label{eq:37}
\end{equation}
and for the decay $\Lambda_{b}\rightarrow\Sigma^{-}D_{s}^{+}$, from
the Eqs. (\ref{eq:30}, \ref{eq:31}, \ref{eq:34}) and Eq. (\ref{eq:35})

\begin{equation}
B=B_{pole}=\left[\frac{G_{F}}{\sqrt{2}}V_{cb}V_{ud}\right]\left(\frac{m_{\Lambda_{b}^{+}}+m_{\Sigma_{b}^{-}}}{m_{\Sigma_{b}^{-}}-m_{\Sigma^{-}}}\right)\sqrt{6}d^{\prime}\frac{\left(-\sqrt{\frac{1}{3}}\right)}{f_{D_{s}}}.\label{eq:38}
\end{equation}

Using, $f_{\pi}=0.131\:\text{GeV}$, $f_{K}=0.161\:\text{GeV}$, and
$f_{D_{s}}=0.257\,\text{GeV}$

\begin{equation}
B_{pole}\approx1.7\times10^{-7},\label{eq:39}
\end{equation}
for the decay $\Lambda_{b}\rightarrow\Lambda_{c}^{+}\pi^{-}$ and
for the decay $\Lambda_{b}\rightarrow\Xi_{c}^{0}K^{0},$

\begin{equation}
B=B_{pole}\approx1.04\times10^{-7}\rightarrow1.8\times10^{-7}\label{eq:40}
\end{equation}
after multiplying by a factor $1.74$ to take into account the $SU(3)$
symmetry breaking. Similarly, for the decay $\Lambda_{b}\rightarrow\Sigma^{-}D_{s}^{+}$

\begin{equation}
B=B_{pole}\approx6.8\times10^{-8}.\label{eq:41}
\end{equation}

For the decays \cite{key-6}

\begin{equation}
\mathcal{B}_{b}^{\prime}\left(p^{\prime}\right)\rightarrow\mathcal{B}\left(p\right)+\mathcal{P}\left(k\right),\label{eq:42}
\end{equation}
the parity-conserving ($p-$wave) decay rate is given by 

\begin{equation}
\Gamma=\frac{k}{4\pi m^{\prime}}\left(p_{0}-m\right)\left|B\right|^{2}.\label{eq:43}
\end{equation}
Hence from Eqs. (\ref{eq:40}, \ref{eq:43});

\begin{equation}
B_{r}\left(\Lambda_{b}\rightarrow\Xi_{c}^{0}K^{0}\right)\approx2.0\times10^{-3}.\label{eq:44}
\end{equation}
and from Eqs. (\ref{eq:41}, \ref{eq:43})

\begin{equation}
B_{r}\left(\Lambda_{b}\rightarrow\Sigma^{-}D_{s}^{+}\right)\approx4.5\times10^{-4}.\label{eq:45}
\end{equation}

To Conclude: in this paper, the formalism of developed in \cite{key-1}
is extended to the bottom baryon decays

\[
\Lambda_{b}^{0}\rightarrow\Sigma_{c}^{*+}\pi^{-},\Xi_{c}^{*0}K^{0};\Lambda_{b}^{0}\rightarrow\Delta^{0}D^{0},\Sigma^{*-}D_{s}^{+}
\]

The most interesting decays are $\Lambda_{b}\rightarrow\Sigma_{c}^{*+}\pi^{-}\rightarrow\Lambda_{c}^{+}\pi^{0}\pi^{-}$
and $\Lambda_{b}\rightarrow\Delta^{0}D^{0}\rightarrow p\pi^{-}D^{0}$.
In another context, decays $\Lambda_{b}\rightarrow\Xi_{c}^{0}K^{0}$
and $\Lambda_{b}\rightarrow\Sigma^{-}D_{s}^{+}$ for which the baryon
poles contribute to the parity-conserving ($p-$wave) amplitude are
analyzed .

Experimental data is not available to check out results. These results
may be of interest to experimental physicists investigating the bottom
baryon decays. The experimental verification of our results will give
a boost to the formalism developed in \cite{key-1}, based on \cite{key-2}
and elaborated in \cite{key-3}.


\begin{thebibliography}{1}
\bibitem{key-1}Fayyazuddin and Riazuddin, Phys. Rev. \textbf{D55},
255 (1997), \textbf{56} 53(E) (1997).

\bibitem{key-2}Riazuddin and Fayyazudin, Phys. Rev. \textbf{D18}
(1978); \textbf{19} 1630 (E) (1978).

\bibitem{key-3}Fayyazuddin and Riazuddin: A Modern scientific, Singapore,
(2011) Chapter 10, section 10.5.3.

\bibitem{key-4}Fayyazuddin, Phys. Rev. \textbf{D 95}, 053008 (2017).

\bibitem{key-5}A. DeRujula, H. Georgi, and Glashow, Phys. Rev. \textbf{D12},
147 (1975).

\bibitem{key-6}See for example, \cite{key-3}, Section 10.5.1.
\end{thebibliography}
\end{document}